%% file: main.tex
\documentclass[10pt,conference]{IEEEtran}
\IEEEoverridecommandlockouts
\usepackage{amsmath,amssymb,amsfonts}
\usepackage{amsmath}
\usepackage{algorithm}
\usepackage{graphicx}
\usepackage{siunitx}
\usepackage{diagbox}
\usepackage[noend]{algpseudocode}
\usepackage{graphicx}
\usepackage{textcomp}
\usepackage{xcolor}
\usepackage{booktabs}
\usepackage{bm}
\usepackage{pifont}
\usepackage{dblfloatfix}
\usepackage[hidelinks]{hyperref}
\usepackage{cleveref}
\usepackage[utf8]{inputenc}
\usepackage{multirow}
\usepackage{textcomp}
\usepackage{comment}
\usepackage{tcolorbox}
\usepackage{balance}
\usepackage[labelformat=simple]{subcaption}

\newcommand{\cmark}{\checkmark}
\newcommand{\xmark}{\ding{55}}

\usepackage{fixltx2e}
\usepackage[
backend=biber
,style=ieee
,minbibnames=1
,maxbibnames=99
,maxcitenames=1
,mincitenames=1
,eprint=false
,doi=false
,isbn=false
,url=false
]{biblatex}
\bibliography{ref} 
\AtBeginBibliography{\footnotesize}

\usepackage{xcolor}

\newcommand{\ad}[1]{\textcolor{blue}{#1}}

\begin{document}

\title{A Replication Study on Predicting Metamorphic Relations at Unit Testing Level}

\author{\IEEEauthorblockN{
	Alejandra Duque-Torres\IEEEauthorrefmark{2},
	Dietmar Pfahl\IEEEauthorrefmark{2}, 
    Rudolf Ramler\IEEEauthorrefmark{3}, and
    Claus Klammer\IEEEauthorrefmark{3}}
			
	\IEEEauthorblockA{%
	\IEEEauthorrefmark{2}\textit{Institute of Computer Science  }, \textit{University of Tartu}, Tartu, Estonia \\
	E-mail: \{duquet, dietmar.pfahl\}@ut.ee}	
	\IEEEauthorblockA{
	\IEEEauthorrefmark{3}\textit{Software Competence Center Hagenberg (SCCH) GmbH}, Hagenberg, Austria \\
	E-mail:  \{rudolf.ramler, claus.klammer\}@scch.at}
}

\maketitle

\begin{abstract}
Metamorphic Testing (MT) addresses the test oracle problem by examining the relations between inputs and outputs of test executions. Such relations are known as Metamorphic Relations (MRs). In current practice, identifying and selecting suitable MRs is usually a challenging manual task, requiring a thorough grasp of the SUT and its application domain. Thus, Kanewala et al. proposed the Predicting Metamorphic Relations (PMR) approach to automatically suggest MRs from a list of six pre-defined MRs for testing newly developed methods. PMR is based on a classification model trained on features extracted from the control-flow graph (CFG) of 100 Java methods. In our replication study, we explore the generalizability of PMR. First, since not all details necessary for a replication are provided, we rebuild the entire preprocessing and training pipeline and repeat the original study in a close replication to verify the reported results and establish the basis for further experiments. Second, we perform a conceptual replication to explore the reusability of the PMR model trained on CFGs from Java methods in the first step for functionally identical methods implemented in Python and C++. Finally, we retrain the model on the CFGs from the Python and C++ methods to investigate the dependence on programming language and implementation details. We were able to successfully replicate the original study achieving comparable results for the Java methods set. However, the prediction performance of the Java-based classifiers significantly decreases when applied to functionally equivalent Python and C++ methods despite using only CFG features to abstract from language details. Since the performance improved again when the classifiers were retrained on the CFGs of the methods written in Python and C++, we conclude that the PMR approach can be generalized, but only when classifiers are developed starting from code artefacts in the used programming language.
\end{abstract}


\section{Introduction}
\label{sec:intro}

\textit{Metamorphic Testing} (MT) is a software testing approach proposed by \citeauthor{chen2020metamorphic}~\cite{chen2020metamorphic} to alleviate the test oracle problem. A test oracle is a mechanism for detecting whether or not the outputs of a program are correct \cite{QuASoQ,6963470}. The oracle problem arises when the SUT lacks an oracle or when developing one to verify computed outputs is practically impossible \cite{6963470}. MT differs from traditional testing approaches in that it examines the relations between input-output pairs of consecutive SUT executions rather than the outputs of individual SUT executions \cite{chen2020metamorphic}. The relations between SUT inputs and outputs are known as \textit{Metamorphic Relations} (MRs). MRs define how the outputs should vary in response to a certain change in the input \cite{6613484,8493260}. In this way, testers may test the SUT indirectly by looking at whether the inputs and outputs satisfy the MRs. 
If an MR is violated for certain test cases, then there is a high probability that there is a fault in the SUT \cite{6613484}. The most challenging task facing MT is determining suitable MRs for a particular SUT. In current practice, MRs are detected manually and require an in-depth understanding of the SUT and problem domain. As a result, the identification and selection of high-quality MRs are recognised as a big challenge. 

Recently, an approach supporting unit testing at the method level was proposed by \citeauthor{PMR3}, published in STVR~\cite{PMR3} and at ISSRE~\cite{PMR1}, to \textit{``predict whether a certain method exhibits a particular MR or not"}. The idea behind their Predicting Metamorphic Relations (PMR) approach is to build a model that predicts whether a method in a newly developed SUT can be tested using a specific MR. The PMR approach is based on a pre-defined set of six MRs and a classifier trained on the control-flow graph (CFG) extracted from a pool of sample methods. While the original study by \citeauthor{PMR3} showed encouraging results, it was performed on a dataset covering methods implemented in one programming language (Java) only. In order to see whether and how the PMR approach could be transferred to other programming languages, given that using CFGs supports abstracting from language and implementation details, we decided to perform further research on this.

Replication studies are required to validate experimental results achieved in prior research. They are an essential part of empirical software engineering as they prove that the observations obtained can hold (or not) under different situations. There are several forms of replication \cite{SHEPPERD2018120, GOMEZ20141033}. An \textit{exact} replication aims to replicate the experiments as closely as possible to the initial procedures. This demonstrates that uncontrolled random variables did not drive the initial results. In a \textit{conceptual} replication, one or more dimensions can be modified to see how well the results hold up. It is important to note that \textit{exact} and \textit{conceptual} replications go by the names of \textit{repetition} and \textit{reproduction}, respectively, in the literature \cite{GOMEZ20141033}. 

In this paper, we present a conceptual replication of the study of \citeauthor{PMR3}~\cite{PMR3}. We follow \ad{the guidelines suggested by \citeauthor{carver2010towards}~\cite{carver2010towards}} and the ACM guidelines on reproducibility (different team, different experimental setup)~\cite{ACM2}: \textit{“The measurement can be obtained with stated precision by a different team, a different measuring system, in a different location on multiple trials. For computational experiments, this means that an independent group can obtain the same or similar result using artefacts, which they develop completely independently”}. We decided to replicate the original study in three steps. In each step, we use the same set of pre-defined MRs used in the original study. 

In the first step, we rebuilt the entire pipeline for extracting CFG information from source code to training and testing classification models in order to repeat the process described in the original study by \citeauthor{PMR3}. The authors of the original study provided the extracted CFG information for replicating their work but not the Java source code of the analyzed methods, which we retrieved from the corresponding project repositories on GitHub. We also had to develop the classification models that we then used for predicting MRs, because also they were not shared by the authors of the original paper. We conducted the first step to check whether we can re-create the classifiers with as good quality as in the original study and to prepare for the next steps by building our own classifiers based on features extracted from the CFGs derived from the source code of methods.

The next two steps of our study aim at exploring the transferability and generalizability of the PMR method. In the second step, we checked whether classifiers generated from Java code perform equally well when applied to Python and C++ code. Therefore, we created two datasets, one comprising $100$ methods in Python and other one comprising $100$ methods in C++. Both sets of methods implemented the exact same functionality as the $100$ Java methods used in the first step. We did this to guarantee that the MRs taken from the original study would apply in the same way to the Python and C++ methods as they did in the original study using Java. To check the generalizability of the PMR method to other programming languages we then also developed individual classifiers for each programming language (Python, C++) starting out from code in the same programming language to which the classifier will be applied during evaluation.


The results of our study indicate that PMR classifiers generated based on artefacts implemented in one programming language do not perform well when applied to artefacts implemented in a different programming language, even though the classifiers are based on CFGs to abstract from programming language and implementation details, the implemented functionality is exactly identical, and the set of MRs remains unchanged. On the other hand, it seems to be possible to generalize the PMR method in the sense that it can be applied with good performance on code implemented in a different programming language as long as the PMR classifiers are redeveloped based on code implemented in the same programming language. 



\section{Related Work}
\label{sec:RelWork}

MT has been demonstrated to be an effective technique for testing in a variety of application domains, \textit{e.g.,} autonomous driving \cite{zhang2018deeproad,zhou2019metamorphic}, cloud and networking systems \cite{canizares2020mt,9477667}, bioinformatic software \cite{10.1145/3193977.3193981, shahri2019metamorphic}, scientific software \cite{peng2021contextual,8533366}. However, the efficacy of MT heavily relies on the specific MRs employed. Some research has been done on how to choose “good" MRs. \citeauthor{chen2004case}~\cite{chen2004case} examined several MRs discovered for shortest path and critical path programs, attempting to determine MRs that are useful. \citeauthor{6319226}~\cite{6319226} introduced the Composition of MRs (CMRs) technique for constructing new MRs by mixing multiple existing ones. \citeauthor{10.1145/2642937.2642994}~\cite{10.1145/2642937.2642994} proposed a method in which an algorithm searches for MRs expressed as linear or quadratic equations. \citeauthor{CHEN2016177}~\cite{CHEN2016177} developed METRIC, a specification-based technique and related tool for identifying MRs based on the category-choice framework. They also expanded METRIC into METRIC+ by integrating the information acquired from the output domain \cite{8807231}. To our knowledge, \citeauthor{PMR3}\cite{PMR1, PMR3}, were the first to show that, in previously unseen methods, MRs can be predicted using ML techniques. They used features obtained from CFGs and a set of predefined MR to train prediction models.

\section{Predicting Metamorphic Relations} 
\label{sec:PMR}

This study is a replication of the approach proposed by \citeauthor{PMR3}~\cite{PMR3} for predicting suitable MRs for the purpose of unit testing. In this section, we first present the PMR procedure proposed by \citeauthor{PMR3} (\Cref{subsec:PMR_procedure}). Then, we present a detailed description of the set of pre-defined MRs used in the original and our study (\Cref{subsec:MetamorphicRlations}) as well as the labelled dataset used in the original study (\Cref{subsec:dataset_PMR}). Finally, we summarise the evaluation results reported in the original study (\Cref{subsec:Eval_Results}).

\subsection{PMR Procedure}
\label{subsec:PMR_procedure}

\begin{figure*}[h!]
	\centering
	\includegraphics[width=\textwidth]{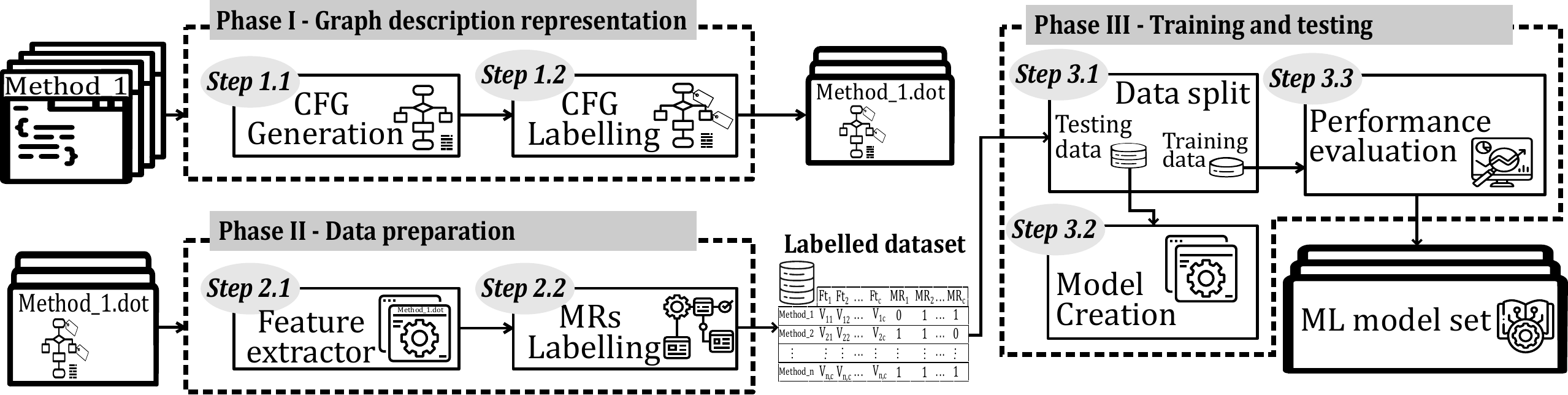}
	\caption{PMR procedure}
	\label{fig:ML-Model}
\end{figure*}

The goal of the PMR approach is to build a model that predicts whether a method in a newly developed SUT can be automatically tested by exploiting one or more MRs contained in a pre-defined set of MRs. \Cref{fig:ML-Model} shows the PMR procedure. The PMR procedure consists of three phases. \textit{Phase I} is responsible for creating a graph description representation derived from a method's control-flow graph (CFG). The output of this phase is a \textit{DOT} file. \textit{Phase II} is in charge of feature extraction from the method's DOT file. Also, each method is labelled with elements from the set of pre-defined MRs. Thus, the output of this phase is a labelled dataset. \textit{Phase III} is in charge of training and evaluating the binary classification models that predict whether a specific MR is applicable to the unit testing of a specific method. Below we describe each phase in detail. 

\subsubsection{{\textbf{Phase I -- Graph description representation}}}
\label{subsubsec:PhaseI}
This Phase starts with the creation of the graph representation from the method's source code. Then a labelled CFG is created by annotating each node in the CFG. The process can be split into the following two steps.

\begin{figure}[ht!]
	\centering
	\includegraphics[width=0.49\textwidth]{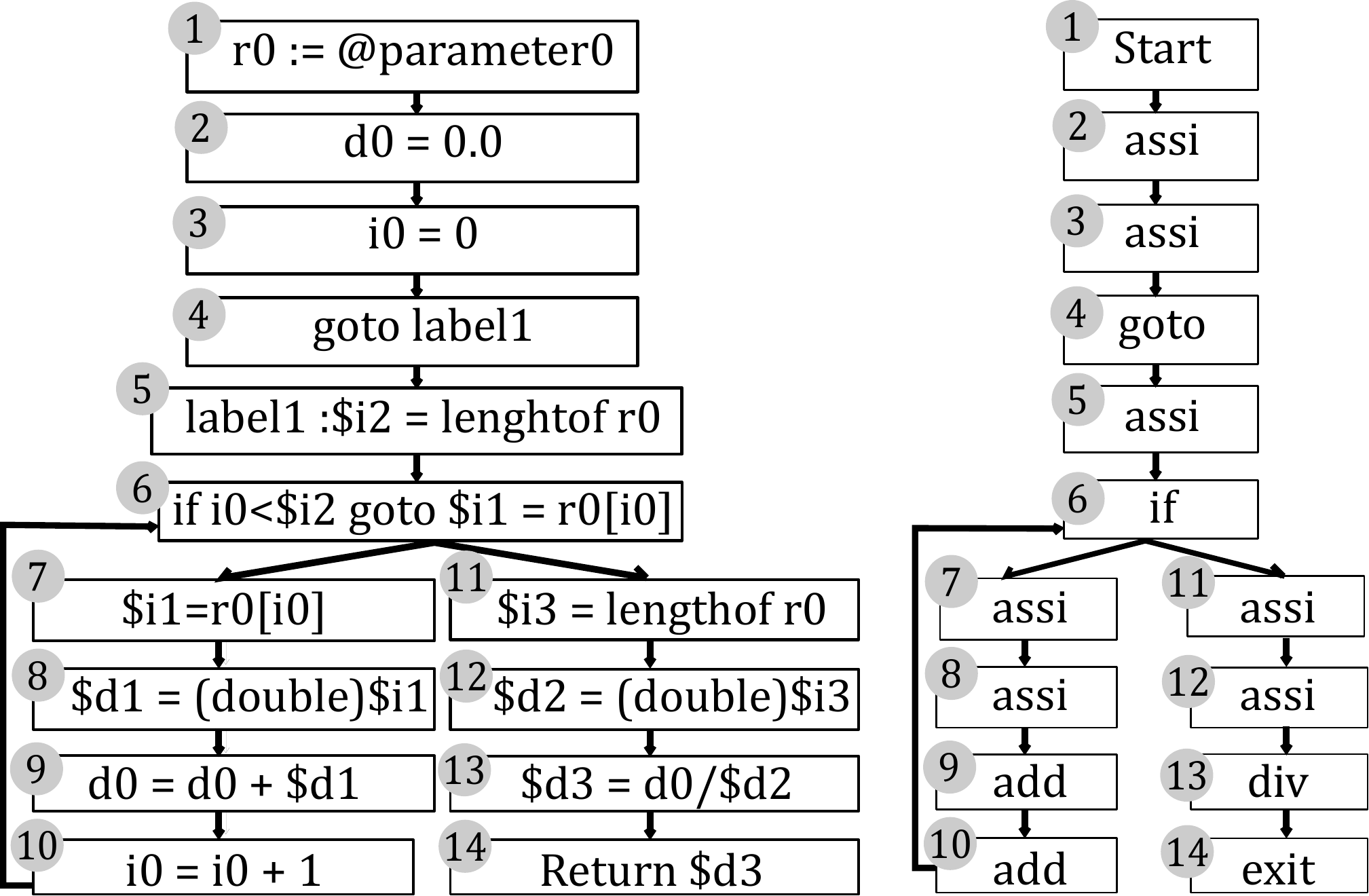}
	\caption{On the left, CFG representation of \Cref{alg:average} using the \textit{soot} framework; on the right, its CFG with annotations}
	\label{fig:CFG_labelled}
\end{figure}

\textit{\textbf{Step 1.1 -- CFG generation:}} This step is responsible for creating the CFG from the method's source code. To get the CFG, \citeauthor{PMR3} use \textit{Soot}\cite{10.1145/1925805.1925818}. \textit{Soot} generates CFG representations in \textit{Jimple} format, a typed 3-address intermediate representation, where each CFG node represents an atomic operation \cite{10.1145/1925805.1925818}. 
The left-hand side of \Cref{fig:CFG_labelled} shows the CFG representation of the \Cref{alg:average} using the \textit{soot} framework. The numbering of the nodes has been done manually, \textit{i.e.,} not with the framework, and serves here only to facilitate a better understanding of the next phase and its steps.

\begin{algorithm}[ht!]
	\caption{Average of an integer array}\label{alg:average}
	\begin{algorithmic}[1]
		\Function{avg}{$int~input[~]$}
		\State \textbf{double} sum = 0;
		\State \textbf{double} average = 0;
		\For{(int i = 0; input.length; i++)}
		\State sum $+=$ input[i];
		\EndFor	
		\State average $=$ sum$/$input.length;
		\State\Return average
		\EndFunction
	\end{algorithmic}
\end{algorithm}

\textit{\textbf{Step 1.2 -- CFG labelling:}} In this step a simplified version of the CFG is created by replacing the specific, code-related information of each node in the CFG by a more general annotation  describing the specific operations and conditional jumps in the original code. The right-hand side of \Cref{fig:CFG_labelled} shows the CFG representation with the node annotations of the \Cref{alg:average}. \Cref{tbl:labels_node} shows examples of annotations that are assigned depending on the node operation. The annotations follow the graph description language,  \textit{i.e.,} the DOT format. 

\subsubsection{\textbf{Phase II -- Data preparation}}
\label{subsubsec:PhaseII}
This phase is in charge of extracting a set of features from the annotated CFGs, \textit{i.e,} from the Phase I output. Also, to each method's annotated CFG zero to six pre-defined MRs are assigned, depending on their suitability. Like Phase I, Phase II consists of two steps. 

\textit{\textbf{Step 2.1 -- Feature extraction:}} \citeauthor{PMR3} propose two approaches for extracting features from CFG representations, features based on nodes and paths, and features based on graph similarity measures. In the former, the node features (hereafter simply NF) follows the form $NO_{n}-d_{in}-d_{out}$, where NO$_{n}$ stands for Node Operation of node n, and $d_{in}$ and $d_{out}$ stand for \textit{in-degree} and \textit{out-degree}, respectively. The number of a specific NF type, \textit{i.e.,}  $NO_{n}-d_{in}-d_{out}$, is tallied and used as the NF value. As an example of NF, let us consider the annotated CFG of \Cref{alg:average}. As the right-hand side of \Cref{fig:CFG_labelled} shows, the annotated CFG of \Cref{alg:average} has fourteen nodes. Among these fourteen nodes, there are seven with the type annotation \textit{assi}, two with type annotation \textit{add}, and five with unique type annotations, \textit{i.e,} \textit{start, goto, if, div} and \textit{exit}. For each node, the $d_{in}$ and $d_{out}$ are calculated, too, to derive the complete NF. For instance, the node \textit{start} (node $1$) will be represented by the NF \textit{start-0-1}, where \textit{start} is $NO_{1}$, $0$ is $d_{in}$ and $1$ is $d_{out}$. Each unique NF is tallied to get the corresponding NF value. For \textit{start-0-1}, the NF value is $1$. \Cref{tbl:node_features} shows the set of NFs and their values extracted from \Cref{alg:average} using its CFG representation.

The feature based on path information (hereafter just PF) refers to the shortest routes from the start node to each node in the graph, as well as the shortest path from each node in the graph to the end node. This feature follows the form $NO_{1}-NO_{2}-NO_{...}-NO_{n}$, where $NO_{n}$, as in the NF, denotes a specific operation statement in node n. The value of each PF is the number of occurrences of each path in the CFG. For instance, let us consider the labelled CFG of \Cref{alg:average}. As \Cref{fig:CFG_labelled} right side shows, the path composed by the nodes the PF \textit{1-2-3-4-5-6-7} and the nodes \textit{1-2-3-4-5-6-11} can be denoted as \textit{start-assi-assi-goto-assi-if-assi}. Therefore, its feature value is 2 since there are two paths represented by one type of PF. \Cref{tbl:path_features} shows the set of PFs and their associated PF values extracted from \Cref{alg:average} using its CFG representation.

The second approach to extract features from CFG is by using \textit{graph similarity measures}. Graph similarity refers to the process of determining the degree of similarity between two or more graphs. In particular, \citeauthor{PMR3} use Random Walk Kernel (RWK) and Graphlet Kernel (GK). RWK is  the most-studied family of graph kernels \cite{10.1007/978-3-540-45167-9_11}. It provides measure the similarity between two or more graphs based on the number of common walks in the graphs.~The concept behind GK is to randomly sample tiny (connected) sub-graphs of size $k$, and using them to compare frequency distributions or to construct graph invariants.

\begin{table}[htbp]
\centering
\caption{Node operations (NO) in the control flow graph and corresponding labels (CL) for the annotation}
{	\label{tbl:labels_node}
	\resizebox{\linewidth}{!} {
	\begin{tabular}{ll|ll|ll|ll}
		\toprule
		\textbf{NO} & \textbf{CL} & \textbf{NO} & \textbf{CL} & \textbf{NO} & \textbf{CL} & \textbf{NO} & \textbf{CL} \\
		\toprule
		$+$ & add  & $-$ & sub & $*$ & mul & $/$ & div \\
		$||$, \textit{$or$} & or & $\&$, \textit{and} & and & \textit{if} & if & $=$ & assi \\
		$==$ & eql & $>=$ & geql & $>$ & gt & $<=$ & leql \\
		$<$  & lt  & $!=$ & neql & $:=$ & start & $\%$ & rem  \\
		\textit{invoke} & fcall & \textit{return} & return & \textit{exit} & exit & \textit{goto} & goto \\
		\bottomrule
	\end{tabular}}}
\end{table}

\textit{\textbf{Step 2.2 -- MR labelling:}} The key idea of PMR is predicting whether a given method is suited for a particular MR by using binary classifiers. PMR uses supervising learning classification algorithms, \textit{i.e.,} a labelled dataset is needed to provide examples for learning. Thus, after \textit{Step 2.1 -- Feature extraction}, the training dataset is created by manually labelling each method with applicable MRs. Depending on whether a specific MR does or does not satisfy the method, the method is labelled with 1 or 0 for this MR, respectively.

\subsubsection{\textbf{Phase III -- Training and testing}}
\label{subsubsec:PhaseIII}
This phase involves the use of one or more supervised machine learning (ML) algorithms, or a combination of them, to derive knowledge from the data. Three steps needs to be conducted.

\textit{\textbf{Step 3.1 -- Data split:}} This step is responsible for splitting the dataset into two subsets: a training set and a test set. The training set is used to create the prediction model, while the test set is used to evaluate the performance of the created prediction model. 

\begingroup
\setlength{\tabcolsep}{6pt} 
\renewcommand{\arraystretch}{1} 
\begin{table}[bp]
\centering
\caption{Node Features (NF) extracted from \Cref{alg:average} related to the nodes of its CFG representation}
{\tiny
	\label{tbl:node_features}
	\resizebox{0.9\linewidth}{!} {
	\begin{tabular}{lc|lc}
		\toprule
		\textbf{NF} & \textbf{NF value} & \textbf{NF} & \textbf{NF value}\\
		\toprule
		\textit{start-0-1}& 1 & \textit{if-2-2}   & 1 \\
		\textit{assi-1-1} & 7 & \textit{add-1-1}  & 2 \\
		\textit{goto-1-1} & 1 & \textit{div-1-1}  & 1 \\
		\bottomrule
	\end{tabular}}}
\end{table}
\endgroup

\begingroup
\setlength{\tabcolsep}{6pt} 
\renewcommand{\arraystretch}{1} 
\begin{table}[tp]
\centering
\caption{Path Features (PF) extracted from \Cref{alg:average} related to the paths of its CFG representation}
{
	\label{tbl:path_features}
	\resizebox{\linewidth}{!} {
	\begin{tabular}{lc}
		\toprule
		\textbf{PF} & \textbf{PF value} \\
		\toprule
		\multicolumn{2}{c}{\textbf{Shortest path from the start node to each node}} \\
		\midrule
		\textit{start} & 1 \\
		\textit{start-assi} & 1  \\
		\textit{start-assi-assi} & 1 \\
		\textit{start-assi-assi-goto} & 1 \\
		\textit{start-assi-assi-goto-assi} & 1 \\
		\textit{start-assi-assi-goto-assi-if} & 1 \\
		\textit{start-assi-assi-goto-assi-if-assi} & 2 \\
		\textit{start-assi-assi-goto-assi-if-assi-assi} & 2 \\
		\textit{start-assi-assi-goto-assi-if-assi-assi-add} & 1 \\
		\textit{start-assi-assi-goto-assi-if-assi-assi-div} & 1 \\
		\textit{start-assi-assi-goto-assi-if-assi-assi-add-add} & 1 \\
		\textit{start-assi-assi-goto-assi-if-assi-assi-div-exit} & 1 \\
		\midrule
		\multicolumn{2}{c}{\textbf{Shortest path from  each  node to the end node}}\\
		\midrule
		\textit{assi-assi-goto-assi-if-assi-assi-div-exit} & 1 \\
		\textit{assi-goto-assi-if-assi-assi-div-exit} & 1 \\
		\textit{goto-assi-if-assi-assi-div-exit} & 1 \\
		\textit{assi-if-assi-assi-div-exit} & 1 \\
		\textit{if-assi-assi-div-exit} & 1 \\
		\textit{assi-assi-div-exit} & 1 \\
		\textit{assi-div-exit} & 1 \\
		\textit{div-exit} & 1 \\
        \textit{exit} & 1 \\
        \textit{assi-assi-add-add-if-assi-assi-div-exit} & 1 \\
        \textit{assi-add-add-if-assi-assi-div-exit} & 1 \\
        \textit{add-add-if-assi-assi-div-exit} & 1 \\
        \textit{add-if-assi-assi-div-exit} & 1 \\
		\bottomrule
	\end{tabular}}}
\end{table}
\endgroup

\textbf{\textit{Step 3.2 -- Model creation}} refers to the process of building prediction models. Choosing a good modelling technique is vital for the training and prediction stage in any ML application, including the PMR approach. \citeauthor{PMR3} get the best results using the Support Vector Machine (SVM) technique.

\textit{\textbf{Step 3.3 -- Performance evaluation:}} This step measures the performance of the created prediction models. Performance measures are derived from the \emph{Confusion Matrix}. Let $A$ denote a classification output in which a specific $MR_{n}$ satisfies the method $m$,  and let $A^{\prime}$ denote a classification output in which a specific $MR_{n}$ does not satisfy the method $m$, then $A$ can be seen as the \textit{positive class} and $A^{\prime}$ as the \textit{negative class}. Using this notation, each standard performance measure is expressed as a function of the counts of elements in the \emph{Confusion Matrix} defined as follows:

\begin{LaTeXdescription}
	\item \emph{True Positive (TP):} The actual MR of a method was $A$ and the predicted was $A$. This represents a successful prediction.
	\item \emph{True Negative (TN):} The actual MR of a method was $A^{\prime}$, the predicted was $A^{\prime}$. This represents a successful prediction.
	\item \emph{False Positive (FP):} The actual MR was $A^{\prime}$ and the predicted was $A$. This represents an unsuccessful prediction.
	\item \emph{False Negative (FN):} The actual MR was $A$ and the predicted was $A^{\prime}$. This represents an unsuccessful prediction.
\end{LaTeXdescription}

\noindent \emph{Accuracy} is the ratio of successful predictions made to both classes and expressed as:
\vspace{-0.5ex}
\begin{equation}
Accuracy = \frac{TP + TN}{TP + TN + FP + FN} 
\label{eq:rw_accuracy}
\end{equation}

\noindent Precision (or positive predictive value) is the ratio of correct predictions made for class $A$ and is shown in \Cref{eq:rw_precision}:
\vspace{-0.5ex}
\begin{equation}
Precision = \frac{TP}{TP + FP}
\label{eq:rw_precision}
\end{equation}

Recall (or true positive rate, or sensitivity) is the ratio of successful predictions made to cases of class $A$
\vspace{-0.5ex}
\begin{equation}
Recall = \frac{TP}{TP + FN}
\label{eq:rw_dr}
\end{equation}

\noindent The \emph{f-measure} statistic (or F1 score) is the harmonic mean of precision and recall:
\vspace{-0.5ex}
\begin{equation}
\text{\normalfont f-measure} = 2 \times \frac{Recall \times Precision}{Recall + Precision} 
\label{eq:rw_fm}
\end{equation}

In addition to the aforementioned performance measures, the \emph{Area Under Curve} (AUC) and the \textit{Balanced Success Rate} (BSR) measures are also widely used. The AUC is the area under the curve that plots the False Positive Rate (FPR) against the True Positive Rate (TPR) at different points in $[0, 1]$. In binary classification problems, the BSR measure is calculated as the average of recall obtained on each class \cite{Ben-Hur2010}.
 
\subsection{Metamorphic Relations}
\label{subsec:MetamorphicRlations}
In the original study, \citeauthor{PMR3} use six MRs that had been suggested previously in other studies \cite{PMR1, PMR2, PMR4, PMR5, PMR6}. Below we describe each MR in detail.  The acronyms \textit{stc} and \textit{ftc} stand for \textit{source test case} and \textit{follow-up test case}, respectively. The input of the stc is an ordered set of non-negative numbers:
\begin{center}
    $Input_{stc} = {X_i, ~..., X_n}$ where $X_i\geq0$, $0 \leq i \leq n$
\end{center}

The outputs of the source and follow-up test cases are written as $Output_{stc}(X)$ and $Output_{ftc}(Y)$, respectively. 

The MRs based on these inputs and outputs are:

\vspace{1ex}
\begin{LaTeXdescription}
    \item \textit{MR$_1$: \textbf{“Addition”}} (\textbf{ADD}). To get the ftc input, add a positive constant “C” to each element of the stc input, \textit{i.e.}, 
    \begin{center}
        $Input_{ftc}={X_1 + C,~X_2 + C,~X_3 + C..., X_n + C}$,
    \end{center}
    Then the following output-relation must hold:
    \begin{center}
        $Output_{ftc} (Y) \geq Output_{stc}(X)$
    \end{center}
    
    \item \textit{MR$_2$: \textbf{“Multiplication”}} (\textbf{MUL}). To get the ftc input, multiply each stc input element with a positive constant “C”, \textit{i.e.}, 
    \begin{center}
        $Input_{ftc}={X_1 * C,~X_2 * C,~...,~X_n * C}$
    \end{center}
    Then the following output-relation must hold:
    \begin{center}
        $Output_{ftc} (Y) \geq Output_{stc}(X)$
    \end{center}
    
    \item \textit{MR$_3$:\textbf{“Permutation”}} (\textbf{PER}). To get the ftc input, randomly permute the stc input elements, \textit{e.g.}, like
    \begin{center}
        $Input_{ftc}={X_3,~X_1,~X_n,~...,~X_2}$
    \end{center}
    Then the following output-relation must hold:
    \begin{center}
        $Output_{ftc} (Y) = Output_{stc}(X)$
    \end{center}

    \item \textit{MR$_4$: \textbf{“Inclusive”}} (\textbf{INC}). To get the ftc input, include a new element “$X_{n + 1} \geq 0$” to the stc input, \textit{e.g.}, like
    \begin{center}
        $Input_{ftc} ={X_1,~X_2,~X_3,~...,~X_n,~X_{n + 1}}$
    \end{center}
    Then the following output-relation must hold:
    \begin{center}
        $Output_{ftc} (Y) \geq Output_{stc}(X)$    
    \end{center}
    
    \item \textit{MR$_5$: \textbf{“Exclusive”}} (\textbf{EXC}). To get the ftc input, remove an element “$X_{n - 1} \geq 0$” from the stc input, \textit{e.g.}, like
    \begin{center}
        $Input_{ftc} ={X_1,~X_2,~X_3,~...,~X_{n - 1}}$    
    \end{center}
    Then the following output-relation must hold:
    \begin{center}
        $Output_{ftc} (Y) \leq Output_{stc}(X)$    
    \end{center}
    
    \item \textit{MR$_6$: \textbf{“Invertive”}} (\textbf{INV}). To get the ftc input, take the inverse of each stc input element $X_i > 0$, \textit{i.e.},
    \begin{center}
        $Input_{ftc} ={1/X_1,~1/X_2,1/X_3,~..., 1/X_n}$
    \end{center}
    Then the following output-relation must hold:
    \begin{center}
        $Output_{ftc} (Y) \leq Output_{stc}(X)$    
    \end{center}
\end{LaTeXdescription}

\subsection{Dataset}
\label{subsec:dataset_PMR}

In their original study, \citeauthor{PMR3} relied on a code corpus containing 100 Java methods that take numerical inputs and produce numerical outputs. The methods are from the open-source libraries \textit{Colt Project}~\cite{colt}, which is an open-source library written for high-performance scientific and technical computing, \textit{Apache Mahou}~\cite{Mahout}, which is a machine learning library, \textit{Apache Commons Mathematics}~\cite{commons-maths}, which is a Library of mathematics and statistics components, and \textit{Java Collections}~\cite{coll}, which is a framework that provides an architecture to store and manipulate the group of objects. All of these libraries are written in Java. 

To create a training dataset, \citeauthor{PMR3} manually labelled each method with the set of pre-defined MRs in a binary manner, \textit{i.e.,} if $MR_{n}$ matches a method $m$, then this method is assigned the label $1$ for $MR_{n}$, otherwise it is $0$. 

\Cref{tbl:MR_specifications} reports the total number of methods that do and do not match a specific MR. One sees that more than half of the methods match with MRs denoted as ADD, MUL and INV, while approximately one third of the methods match with MRs denoted as PER, INC, and EXC.

\begingroup
\setlength{\tabcolsep}{6pt} 
\renewcommand{\arraystretch}{1} 
\begin{table}[ht!]
\centering
\caption{Total number of methods that match (\cmark) and do not match (\xmark) a specific MR}
{
	\label{tbl:MR_specifications}
	\resizebox{\linewidth}{!} {
	\begin{tabular}{l|l|l|l|l}
		\toprule
		\textbf{MR} & \textbf{Change in the input} & \textbf{Output expected} & \cmark & \xmark \\
		\toprule
	    ADD & Add a positive constant         & Increase or remain constant & $56$ & $44$   \\
		MUL & Multiply by a positive constant & Increase or remain constant & $66$ & $34$  \\
		PER & Permute the components          & Remain constant             & $33$ & $67$  \\
		INC & Add a new element               & Increase or remain constant & $34$ & $66$  \\
		EXC & Remove an element               & Decrease or remain constant & $32$ & $68$  \\
		INV & Take the inverse of each element& Decrease or remain constant & $63$ & $37$  \\
		\bottomrule
		
	\end{tabular}}}
\end{table}
\endgroup

\Cref{tbl:MR_freq} reports how many methods have 0, 1, 2, ... 6 matching MRs and how those MRs are distributed in each case. $20$ out of $100$ methods have no matching MR, and only $9$ out of $100$ methods match with all six MRs simultaneously.


\begingroup
\setlength{\tabcolsep}{6pt} 
\renewcommand{\arraystretch}{1} 
\begin{table}[ht]
\centering
\caption{Total number of methods that have 0, 1, 2, ..., 6 matching MRs and their distributions}
{
	\label{tbl:MR_freq}
	\resizebox{\linewidth}{!} {
	\begin{tabular}{c|c|cccccc}
		\toprule
		\textbf{No. MR$^\nmid$} & \textbf{No. Met$^\perp$} & \textbf{ADD} & \textbf{MUL} & \textbf{PER} & \textbf{INC} & \textbf{EXC} & \textbf{INV} \\
		\toprule
	    0 &	20 & 0 & 0 & 0 & 0 & 0 & 0 \\
        1 &	8  & 2 & 3 & 0 & 2 & 0 & 1 \\
        2 &	7  & 3 & 4 & 2 & 1 & 1 & 3 \\
        3 &	23 & 19& 17& 5 & 5 & 5 & 18\\
        4 &	26 & 16& 26& 16& 10& 10& 26\\
        5 &	7  & 7 & 7 & 1 & 7 & 7 & 6 \\
        6 &	9  & 9 & 9 & 9 & 9 & 9 & 9 \\
		\bottomrule
		\multicolumn{8}{l}{$^\nmid$Number of MRs that may apply to certain method, $^\perp$Number of methods}
	\end{tabular}}}
\end{table}
\endgroup

\input{table_dataset}

\Cref{tbl:dataset_with_MRs} shows the names of all methods used in the study and the library to which they belong. It also shows whether or not a specific MR matches the method. For better readability, we use the symbol \cmark~to denote that a specific MR matches, otherwise, we use the symbol \xmark. In total, $26$ methods stem from the \textit{Colt project}, $8$ are from \textit{Apache Mahout}, $25$ are from \textit{Apache Commons Mathematics}, and $41$ methods are from \textit{Java Collections}. These methods are provided by \citeauthor{PMR3}\footnote{\href{http://www.cs.colostate.edu/saxs/MRpred/functions.tar.gz}{http://www.cs.colostate.edu/saxs/MRpred/functions.tar.gz}} in the form of CFG representation, using the graph description language format DOT, instead of source code. 

\subsection{Achieved Performance in Original Study}
\label{subsec:Eval_Results}
\citeauthor{PMR3} use features based on nodes and paths, as well as features based on graph similarity (RWK and GK), to build 18 binary SVM models, \textit{i.e.,} three models (each using different feature sets) per any of the six specific MRs. They use AUC and BSR to evaluate model performance and consider AUC $>0.80$ to be a good classification performance. Among the eighteen trained SVM models, the most promising one was when RWK was used. The average performance for the six models using RWK was $0.87$ in terms of AUC. 

\section{Replication Methodology}
\label{sec:methodology}
Our goal is to investigate whether the PMR approach of \citeauthor{PMR3} (i) can be replicated when using our own implementation of the pipeline for developing classifiers starting out from Java source code instead of CFG representations, (ii) classifiers trained on Java source code can be transferred to Python and C++ methods that have identical functionality, and (iii) the PMR approach can be applied to Python and C++ code when classifiers are developed from scratch in the target programming languages. Each of these scenarios gives rise to a research question that we answer in our study. 
In all scenarios, we follow the PMR procedure, \Cref{fig:ML-Model}, and we use the same set of six pre-defined MRs used in the original study, \textit{i.e., }\Cref{subsec:MetamorphicRlations}. However, we develop our own pipeline and create new datasets. For performance evaluation, we employ 10-fold stratified cross-validation., \textit{i.e.,} the dataset is randomly partitioned into ten subgroups. The classifier is then built using nine subsets, with the 10th subset being used to evaluate the predictive model's performance. This procedure is done ten times, with each of the ten subgroups being evaluated separately. The ten folds in stratified 10-fold cross-validation are partitioned in such a way that they include about the same proportion of classes as the original data set. The resulting overall performance is measured as the average of the ten cross-validation tests.

Our replication package containing results, scripts, models and datasets is available online\footnote{\href{https://github.com/aduquet/RENE-PredictingMetamorphicRelations}{https://github.com/aduquet/RENE-PredictingMetamorphicRelations}}.

\subsection{Research Questions}
\label{subsec:RQs}
We aim at answering the following research questions
\begin{itemize}
    \item \textbf{RQ$_{1}$:} [Replicability] \textit{~How well do classifiers predict matching MRs for Java methods when using our processing and training pipeline starting from source code?}
    \item \textbf{RQ$_{2}$:} [Transferability] \textit{~How well do classifiers developed on Java code predict matching MRs for functionally equivalent methods implemented in Python and C++?}
    \item \textbf{RQ$_{3}$:} [Generalisability] \textit{~How well do classifiers predict matching methods for Python and C++ methods when developed from source code in the respective target languages?}
\end{itemize}

\subsection{RQ$_{1}$:~How well do classifiers predict matching MRs for Java methods when using our pipeline implementation?}
\label{subsubsec:RQ1}
In RQ1, we investigate the replicability of the PMR approach when starting out directly from Java source code (instead of CFG representations), performing all steps of feature extraction and re-generating the classifiers with a different ML package. In particular, we are interested in checking whether the classifiers developed by us achieve the same performance as published in \cite{PMR3}. Satisfactory results for RQ1 are the pre-requisite for tackling RQ2 and RQ3. 

To compare with the work of \citeauthor{PMR3}, we develop our own artefacts for  \textit{Phase II - Data preparation, Step 2.1 -- Feature extraction}, and all the artefacts needed by \textit{Phase III Training and testing}. Then we compare the results of SVM obtained by \citeauthor{PMR3}~\cite{PMR3} who used \textit{PyML Toolkit}~\cite{PyML} with our results achieved using Python \textit{scikit-learn} library~\cite{scikit-learn} with default parameter settings. 
For the comparison we use two datasets. The first dataset is the one used in the original study by \citeauthor{PMR3}, \textit{i.e.,} the methods from \Cref{tbl:dataset_with_MRs}. This dataset contains the CFG representations of the $100$ Java methods in DOT format. In the following, we call this dataset $DS_{JK}$. The second dataset contains the Java source code of the 100 methods from \Cref{tbl:dataset_with_MRs} as contained in the open-source libraries. We call this dataset $DS_{JV}$.

When using  $DS_{JK}$ we apply the PMR approach from \textit{Phase I - Step 1.2} through \textit{Phase III - Step 3.3}, since this dataset already contains the CFG representation of each method in DOT format. When using  $DS_{JV}$, we apply our entire pipeline implementing the PMR approach, \textit{i.e.,} from \textit{Phase I - Step 1.1} to \textit{Phase III - Step 3.3}. To get the CFG representation in DOT format for the dataset $DS_{JV}$, we use the \textit{soot}\cite{10.1145/1925805.1925818} Java framework, configured so that the output matches the CFGs of the original dataset. Then, we compare the performance of our classifiers against the results published by \citeauthor{PMR3}. In particular, we compare BSR and AUC since these are the measures provided in the original study \cite{PMR3}. In addition, we provide the performance measures detailed in \Cref{subsubsec:PhaseIII} - \textit{Step 3.3 -- Performance evaluation}.

\subsection{RQ$_{2}$:~How well do classifiers developed on Java code predict matching MRs for functionally equivalent methods implemented in Python and C++?} 
\label{subsubsec:RQ2}
In RQ2, we check whether classifiers developed with the PMR approach from Java methods achieve the same performance when applied to methods with identical functionality but implemented in Python or C++. We chose Python and C++ because both are popular and widely used programming languages supporting a broad range of applications~\cite{cass2020top}. 

For this experiment, we created two new datasets containing source code of methods written in Python and in C++. The methods in each dataset are functionally identical to that of the $100$ Java methods described in \Cref{tbl:dataset_with_MRs}. The corresponding method implementations were either retrieved from the \emph{NumPy} package for scientific computing in case of Python, the \emph{Blinz++} high-performance library for scientific computing in case of C++, or they were implemented in Python/C++ by the authors if not present in these libraries. Because the functionality of the Python and C++ methods is equivalent to the functionality of the Java methods, we can assume that exactly the same MRs that match the Java methods match the corresponding Python and C++ methods. The dataset named $DS_{PY}$ contains the Python methods, and the dataset named $DS_{C++}$ contains the C++ methods. To get the graph representation in DOT format, \Cref{subsubsec:PhaseI}: \textit{Phase I - Step CFG generation}, for the methods written in Python, we use the Python package \textit{pycfg}~\cite{pycfgcite}, and for the methods written in C++, we use \textit{Goblint}\cite{Goblint2016,goblintGit}. 

\subsection{RQ$_{3}$:~How well do classifiers predict matching methods for Python and C++ methods when developed from source code in the respective target languages?}
\label{subsubsec:RQ3}

Finally, in RQ3, we check whether the PMR approach works for Python and C++ code similarly well as it does for Java code, if we develop the classifiers for each target language from scratch. Thus, we train SVM models using each dataset, \textit{i.e.,} $DS_{PY}$ and $DS_{C++}$. We compare the performance of the new classifiers against the results obtained in RQ1 and RQ2. 

\section{Results and Discussion}
\label{sec:results}


\input{RQ1_JK-JV}
\input{RQ1}
\input{RQ2_PY-C++}

\subsection{RQ$_1$~How well do classifiers predict matching MRs for Java methods when developed from source code using our pipeline?}
\label{subsec:RQ1_results}
\Cref{tbl:RQ1_JK-JV} shows the performance of our PMR implementation for both Java datasets, $DS_{JK}$ and $DS_{JV}$. Overall, regardless of the feature extraction technique used, the results are fairly close. This can be seen in the Error column, which displays the difference in performance between $DS_{JK}$ and $DS_{JV}$ for each MR. The most negative value is  $-0.104$ (Accuracy of ADD) while the farthest positive value is $0.148$ (BSR of INV). This indicates that the classifiers developed by us are consistent for Java code independent from the starting point of the model development (CFG vs. source code). 

\Cref{tbl:RQ1} shows how the performance of our PMR implementation compares to the performance obtained by \citeauthor{PMR3} in terms of AUC and BSR. As can be seen from the Error column, our results are close to those obtained in the original study. The Error range is [-0.093, 0.061] for AUC and [-0.118, 0.117] for BSR. From combining the results shown in \Cref{tbl:RQ1} with those shown in \Cref{tbl:RQ1_JK-JV} we conclude that our implementation of PMR achieves similar performance as reported in \cite{PMR3} even when starting out from source code.




\vspace{2ex}
\begin{tcolorbox}
With regards to \textit{replicability} (RQ1), our results indicate that we can achieve similar results as \citeauthor{PMR3} when re-implementing the PMR approach no matter whether we start the modelling process from source code or from CFG representations.
\end{tcolorbox}

\input{RQ3_PY-C++}

\subsection{RQ$_2$~How well do classifiers developed on Java code predict matching MRs for functionally equivalent methods implemented in Python and C++?}
\label{subsec:RQ2_results}

\Cref{tbl:RQ2_PY-C++} reports on the performance when using classifiers, developed starting out from the $DS_{JV}$ dataset, to predict matching MRs for methods contained in the $DS_{PY}$ and $DS_{C++}$ datasets. The assumption behind applying a classifier built on Java code to methods that are functionally equivalent but implemented in a different programming language is that the CFG representations from which the features in the SVM models are taken would be similar enough to achieve similar classification performance as when applied to Java methods. 


However, as shown in \Cref{tbl:RQ2_PY-C++}, the performance is low for all performance measures and for both Python and C++. No measure is greater than $0.689$. This result suggests that the representation of the CFGs of the Python and C++ methods to which the feature extraction algorithm is applied are more different from the CFGs of the Java methods than expected. This can be explained due to the language-specific CFG generators that we used as well as differences in the way how the methods (with identical functionality) are implemented in different programming languages.

\vspace{2ex}
\begin{tcolorbox}
With regards to \textit{transferability} (RQ2), our results suggest that classifiers trained on a dataset containing methods in one programming language (Java) have reduced performance when applied to datasets with functionally equivalent methods implemented in a different programming language (Python, C++). Hence, classification models built according to the proposed PMR approach may not be be transferable across languages.
\end{tcolorbox}

\subsection{RQ$_3$~How well do classifiers predict matching methods for Python and C++ methods when developed from source code in the respective target languages?}
\label{subsec:RQ3_results}

\Cref{tbl:RQ3_PY-C++} reports the results of using the PMR approach to develop classifiers separately for each programming language (Python and C++). Comparing \Cref{tbl:RQ2_PY-C++} and \Cref{tbl:RQ3_PY-C++} indicates that the performance improves remarkably when using models that are trained specifically to also consider the implementation characteristics stemming from the different programming languages. 
Even though the performance has improved by developing language specific classifiers, the results for Python and C++ are generally below the results achieved for Java, with the results for C++ being consistently the worst.

\vspace{2ex}
\begin{tcolorbox}
With regards to \textit{generalizability} (RQ3), our results suggest the PMR approach can be applied for different programming languages when the classifiers are re-trained on the specific target language. The slightly lower performance, esp. for C++, needs further exploration of the data and the choice of model parameter settings (tuning). \end{tcolorbox}


\subsection{Threats to Validity}
\label{subsec:threats_validity}
In the context of our study, two types of threats to validity are most relevant: threats to internal and external validity.

To achieve internal validity, we used the same set of methods and of MRs as in \citeauthor{PMR3}~\cite{PMR3}. For the Python and C++ datasets, we carefully checked functional equivalence of the methods with those in the original Java dataset. Given functional equivalence of the methods, we assume that the matching MRs are identical for each of the three chosen programming languages. However, this has not been verified. It is unlikely but possible that some methods have slight differences in the set of matching MRs due to the programming language. Another potential validity threat in our study is that we recreated all steps of the PMR approach using different machine learning libraries with potentially different parameter settings.
However, the performance measures in RQ1 (\Cref{tbl:RQ1_JK-JV}) align well with the results reported in the original study. This suggests that we have understood how to correctly build the classifiers in our replication.

Regarding external validity, our study uses the same methods as in the original study but implemented in different programming languages. For the sake of generalisability, it would have been preferable to include additional methods to overcome any potential bias introduced by the selection of methods in the original study. 
As a consequence, our replication cannot determine the actual scope of the effectiveness of the PMR approach. 

\subsection{Remarks on General Relevance}
\label{subsec:limitation_toAplicability}

When assembling the Python and C++ datasets containing functionally equivalent methods for our replication, we identified the issue that such methods tend to be rare and are usually only found in specific domains such as libraries for mathematical computations. In the original study, a fully labelled dataset was used containing a high number of methods (80\%) with matching MRs. Only 20\% of the methods are not related to any of the supported MRs. How realistic is this distribution? Since the pre-defined set of MRs is rather small and only applies to methods with a very specific signature (mainly functions that take numerical inputs and produce numerical outputs), it is unlikely that one will find an equal share of such methods in real-world applications.
In particular since such methods are often already provided as part of existing, dedicated libraries (e.g., Apache Commons or NumPy).
If, as we assume, the share of matching methods in newly developed real-world application is very small and given that the effort for developing language-specific classifiers is comparably high, the practical relevance of the proposed approach seems to be limited.

Furthermore, the proposed PMR approach uses features extracted from individual methods and it is therefore tied to the level of unit testing. A generalisation of the approach beyond unit testing, e.g., by transferring it to system level testing does not seem possible. 



\section{Conclusion}
\label{sec:conclusion}
We closely as well as conceptually replicated the study of \citeauthor{PMR3}~\cite{PMR3}.
First, we reproduced the PMR approach using our own implementation of the pipeline for feature extraction and training classifiers by starting out from Java source code and creating corresponding CFGs. We showed that our classifiers perform equally well as in the original study indicating a successful replication as basis for further experiments. Second, we checked transferability of classifiers trained on methods implemented in Java to other programming languages (Python and C++). We found that the performance decreases too much to consider this approach feasible.
This is caused by programming language-specific implementation details, despite relying only on features extracted from the abstract CFG representation of the methods. 
Third, we demonstrated that the PMR approach can be generalized. When re-training the classifiers from scratch on Python and C++ source code, the performance we achieved was almost comparable to those from classifiers trained on Java code. 

All artefacts created by us as well as all results are available in a replication package.

\section*{Acknowledgement}
This research was partly funded by the Estonian Center of Excellence in ICT research (EXCITE), the European Regional Development Fund, the IT Academy Programme for ICT Research Development, the Austrian ministries BMVIT and BMDW, the State of Upper Austria under the COMET (Competence Centers for Excellent Technologies) program managed by FFG, and grant PRG1226 of the Estonian Research Council. 
\balance
\printbibliography

\end{document}

%% file: table_dataset.tex
\extrarowheight = 0.25ex

\begin{table*}[!ht]
	\caption{Labelled dataset~\cite{PMR3}: symbol \cmark denotes an MR-method match; symbol \xmark~denotes that there is no match}
	\label{tbl:dataset_with_MRs}
	\centering
	\resizebox{\linewidth}{!} {
	\begin{tabular}{c|l|l|cccccc|c|l|l|cccccc}
		\toprule
		\textbf{ID}& \textbf{Method Name}& \textbf{Library}&\multicolumn{6}{c|} {\textbf{Metamorphic Relation}}&
		\textbf{ID}& \textbf{Method Name}& \textbf{Library}&\multicolumn{6}{c} {\textbf{Metamorphic Relation}}\\
		&&&\textbf{ADD}&\textbf{MUL}&\textbf{PER}&\textbf{INC}&\textbf{EXC}&\textbf{INV}&
		&&
		&\textbf{ADD}&\textbf{MUL}&\textbf{PER}&\textbf{INC}&\textbf{EXC}&\textbf{INV}\\
		\midrule
		1	&	add\_values         & Colle & \cmark& \cmark& \cmark& \cmark& \cmark& \cmark&	
		51	&	find\_median	    & Colle & \cmark& \cmark& \cmark& \xmark& \xmark& \cmark\\
        
        2	&	array\_calc 	    & Colle & \cmark& \cmark& \xmark& \xmark& \xmark& \cmark&	
        52	&	find\_min	        & Colle & \cmark& \cmark& \cmark& \xmark& \xmark& \cmark\\
        
        3	&	array\_copy	        & Colle & \cmark& \cmark& \xmark& \xmark& \xmark& \cmark&	
        53	&	g\_Test	            & Math  & \xmark& \cmark& \xmark& \cmark& \cmark& \cmark\\
        
        4	&	autoCorrelation	    & Colt  & \xmark& \xmark& \xmark& \xmark& \xmark& \xmark&	
        54	&	geometric\_mean	    & Colle & \cmark& \cmark& \cmark& \xmark& \xmark& \cmark\\
        
        5	&	average	            & Colle & \cmark& \cmark& \cmark& \xmark& \xmark& \cmark&	
        55	&	get\_array\_value	& Colle & \cmark& \cmark& \xmark& \cmark& \cmark& \cmark\\
        
        6	&	bi\_SearchFromTo	& Colt & \xmark& \xmark& \xmark& \cmark& \xmark& \xmark&	
        56	&	hamming\_dist	    & Colle& \xmark& \xmark& \xmark& \cmark& \cmark& \cmark\\
        
        7	&	bubble	            & Math & \cmark& \cmark& \cmark& \xmark& \xmark& \cmark&	
        57	&	harmonicMean	    & Colt & \cmark& \cmark& \cmark& \xmark& \xmark& \cmark\\
        
        8	&	cal\_AbsoluteDiff	& Math & \cmark& \cmark& \xmark& \xmark& \xmark& \cmark&	
        58	&	insertion\_sort	    & Colle& \cmark& \cmark& \cmark& \xmark& \xmark& \cmark\\
        
        9	&	cal\_Diff           & Math & \xmark& \xmark& \xmark& \xmark& \xmark& \xmark&	
        59	&	kurtosis	        & Colt & \cmark& \cmark& \cmark& \xmark& \xmark& \xmark\\
        
        10	&	chebyshevDist	    & Maho & \xmark& \cmark& \xmark& \cmark& \cmark& \cmark&	
        60	&	lag 	            & Colt & \xmark& \cmark& \xmark& \xmark& \xmark& \xmark\\
        
        11	&	checkNonNegative	& Math & \xmark& \cmark& \cmark& \cmark& \xmark& \cmark&	
        61	&	manhattanDist	    & Maho & \xmark& \cmark& \xmark& \cmark& \cmark& \cmark\\
        
        12	&	checkPositive	    & Math & \xmark& \cmark& \cmark& \cmark& \xmark& \cmark&	
        62	&	manhattanDist2	    & Colle& \xmark& \cmark& \xmark& \cmark& \cmark& \cmark\\
        
        13	&	check\_equal	    & Colle& \xmark& \xmark& \xmark& \xmark& \xmark& \xmark&	
        63	&	max	                & Colt & \cmark& \cmark& \cmark& \cmark& \cmark& \cmark\\
        
        14	&	check\_eq\_tolerance& Colle& \xmark& \xmark& \xmark& \xmark& \xmark& \xmark&	
        64	&	meanDeviation	    & Colt & \cmark& \xmark& \cmark& \xmark& \xmark& \xmark\\
        
        15	&	chiSquare	        & Math & \xmark& \cmark& \xmark& \xmark& \xmark& \cmark&	
        65	&	mean\_Diff	        & Math & \xmark& \xmark& \xmark& \xmark& \xmark& \xmark\\
        
        16	&	clip	            & Colle& \xmark& \xmark& \xmark& \xmark& \xmark& \xmark&	
        66	&	mean\_abs\_error	& Colle& \xmark& \cmark& \xmark& \xmark& \xmark& \cmark\\
        
        17	&	cnt\_zeroes	        & Colle& \xmark& \xmark& \cmark& \cmark& \cmark& \xmark&	
        67	&	min	                & Colt & \cmark& \cmark& \cmark& \xmark& \xmark& \cmark\\
        
        18	&	canberraDist	    & Math & \xmark& \xmark& \xmark& \cmark& \cmark& \cmark&	
        68	&	partition	        & Math & \xmark& \xmark& \xmark& \xmark& \xmark& \xmark\\
        
        19	&	cal\_DividedDiff	& Math & \cmark& \xmark& \xmark& \xmark& \xmark& \xmark&	
        69	&	polevl	            & Colt & \cmark& \cmark& \xmark& \cmark& \cmark& \cmark\\
        
        20	&	cosineDist	        & Maho & \xmark& \cmark& \xmark& \xmark& \xmark& \xmark&	
        70	&	pooledMean	        & Colt & \cmark& \cmark& \cmark& \xmark& \xmark& \cmark\\
        
        21	&	count\_k            & Colle& \xmark& \xmark& \cmark& \cmark& \cmark& \xmark&	
        71	&	pooledVariance	    & Colt & \cmark& \cmark& \cmark& \xmark& \xmark& \cmark\\
        
        22	&	count\_non\_zeroes	& Colle& \cmark& \cmark& \cmark& \cmark& \cmark& \cmark&	
        72	&	power	            & Colt & \xmark& \cmark& \xmark& \xmark& \xmark& \cmark\\
        
        23	&	covariance	        & Colt & \cmark& \xmark& \xmark& \xmark& \xmark& \xmark&	
        73	&	product	            & Colt & \cmark& \cmark& \cmark& \cmark& \cmark& \cmark\\
        
        24	&	dec	                & Maho & \xmark& \xmark& \xmark& \xmark& \xmark& \xmark&	
        74	&	quantile	        & Colt & \cmark& \cmark& \xmark& \xmark& \cmark& \cmark\\
        
        25	&	dec\_array	        & Colle& \cmark& \cmark& \xmark& \xmark& \xmark& \cmark&	
        75	&	reverse	            & Colle& \cmark& \cmark& \xmark& \xmark& \xmark& \cmark\\
        
        26	&	Dist                & Math & \xmark& \cmark& \xmark& \cmark& \cmark& \cmark&	
        76	&	safeNorm	        & Colle& \cmark& \cmark& \cmark& \cmark& \cmark& \cmark\\
        
        27	&	DistInf	            & Math & \xmark& \cmark& \xmark& \cmark& \cmark& \cmark&	
        77	&	sampleKurtosis	    & Math & \cmark& \xmark& \cmark& \xmark& \xmark& \cmark\\
        
        28	&	dot\_product	    & Colle& \cmark& \cmark& \xmark& \cmark& \cmark& \cmark&	
        78	&	sampleSkew	        & Colt & \cmark& \xmark& \cmark& \xmark& \xmark& \xmark\\
        
        29	&	durbinWatson	    & Colt & \xmark& \cmark& \xmark& \xmark& \xmark& \xmark&	
        79	&	sampleVariance	    & Colt & \cmark& \cmark& \cmark& \xmark& \xmark& \cmark\\
        
        30	&	ebeAdd	            & Math & \cmark& \cmark& \xmark& \xmark& \xmark& \cmark&	
        80	&	sampleWeightedVar	& Colt & \xmark& \xmark& \xmark& \xmark& \xmark& \cmark\\
        
        31	&	ebeDivide	        & Math & \xmark& \xmark& \xmark& \xmark& \xmark& \xmark&	
        81	&	scale	            & Colt & \cmark& \cmark& \xmark& \xmark& \xmark& \cmark\\
        
        32	&	ebeMultiply	        & Math & \cmark& \cmark& \xmark& \xmark& \xmark& \cmark&	
        82	&	s\_add	            & Maho & \cmark& \xmark& \xmark& \cmark& \cmark& \xmark\\
        
        33	&	ebeSubtract	        & Math & \xmark& \xmark& \xmark& \xmark& \xmark& \xmark&	
        83	&	selection\_sort	    & Colle& \cmark& \cmark& \cmark& \xmark& \xmark& \cmark\\
        
        34	&	elemtWise\_equal	& Colle& \xmark& \xmark& \xmark& \xmark& \xmark& \xmark&	
        84	&	sequential\_search	& Colle& \xmark& \xmark& \xmark& \cmark& \cmark& \xmark\\
        
        35	&	elemtWise\_max	    & Colle& \cmark& \cmark& \xmark& \xmark& \xmark& \cmark&	
        85	&	set\_min\_val	    & Colle& \cmark& \cmark& \xmark& \xmark& \xmark& \cmark\\
        
        36	&	elemtWise\_min	    & Colle& \cmark& \cmark& \xmark& \xmark& \xmark& \cmark&	
        86	&	shell\_sort	        & Colle& \cmark& \cmark& \cmark& \xmark& \xmark& \cmark\\
        
        37	&	elemtWise\_not\_eq  & Colle& \xmark& \xmark& \xmark& \xmark& \xmark& \xmark&	
        87	&	skew	            & Colle& \cmark& \cmark& \cmark& \xmark& \xmark& \xmark\\
        
        38	&	entropy	            & Math & \cmark& \cmark& \cmark& \cmark& \cmark& \xmark&	
        88	&	square          	& Colle& \cmark& \cmark& \xmark& \xmark& \xmark& \cmark\\
        
        39	&	equals	            & Math & \xmark& \xmark& \xmark& \xmark& \xmark& \xmark&	
        89	&	standardize	        & Colle& \cmark& \cmark& \xmark& \xmark& \xmark& \xmark\\
        
        40	&	errorRate           & Maho & \xmark& \xmark& \xmark& \xmark& \xmark& \xmark&	
        90	&	sum	                & Maho & \cmark& \cmark& \cmark& \cmark& \cmark& \cmark\\
        
        41	&	euc\_Dist   	    & Math & \xmark& \cmark& \xmark& \cmark& \cmark& \cmark&	
        91	&	sumOfLogarithms	    & Colle& \cmark& \cmark& \cmark& \cmark& \cmark& \cmark\\
        
        42	&	evaluateHoners	    & Math & \cmark& \cmark& \xmark& \cmark& \cmark& \cmark&	
        92	&	sum\_Power\_Deviat	& Colt & \xmark& \xmark& \xmark& \xmark& \xmark& \xmark\\
        
        43	&	eval\_Internal	    & Math & \xmark& \xmark& \xmark& \xmark& \xmark& \xmark&	
        93	&	sum\_labeled	    & Colt & \xmark& \xmark& \xmark& \xmark& \xmark& \xmark\\
        
        44	&	evalNewton	        & Math & \xmark& \xmark& \xmark& \cmark& \xmark& \xmark&	
        94	&	tanimotoDist	    & Maho & \xmark& \xmark& \xmark& \xmark& \xmark& \xmark\\
        
        45	&	evalWeightedProd    & Math & \cmark& \cmark& \xmark& \cmark& \cmark& \cmark&	
        95	&	variance	        & Colle& \cmark& \cmark& \cmark& \xmark& \xmark& \cmark\\
        
        46	&	find\_diff	        & Colle& \xmark& \xmark& \xmark& \xmark& \xmark& \xmark&	
        96	&	var\_Difference	    & Colt & \cmark& \cmark& \xmark& \xmark& \xmark& \cmark\\
        
        47	&	find\_euc\_Dist	    & Colle& \xmark& \cmark& \xmark& \cmark& \cmark& \cmark&	
        97	&	weightedMean	    & Colt & \cmark& \cmark& \xmark& \xmark& \xmark& \cmark\\
        
        48	&	find\_magnitude	    & Colle& \cmark& \cmark& \cmark& \cmark& \cmark& \cmark&	
        98	&	weightedRMS	        & Colt & \xmark& \xmark& \xmark& \xmark& \xmark& \xmark\\
        
        49	&	find\_max	        & Colle& \cmark& \cmark& \cmark& \cmark& \cmark& \cmark&	
        99	&	weighted\_average	& Colle& \cmark& \cmark& \xmark& \xmark& \xmark& \cmark\\
        
        50	&	find\_max2	        & Colle& \cmark& \cmark& \xmark& \cmark& \cmark& \cmark&	
        100	&	winsorizedMean	    & Colt & \cmark& \cmark& \xmark& \xmark& \cmark& \cmark\\
        \bottomrule
        \multicolumn{10}{l}{\textbf{Colle:} \textit{Java Collection}, \textbf{Maho:} \textit{Apache Mahout}, \textbf{Math:} \textit{Apache  Commons Mathematics}}
	\end{tabular}
    }
\end{table*}

%% file: RQ1_JK-JV.tex
\extrarowheight = 0.25ex
\begin{table*}[!ht]
	\caption{PMR performance achieved by our classifiers when starting from DS$_{JV}$ and DS$_{JK}$}
	\label{tbl:RQ1_JK-JV}
	\centering
	\resizebox{\textwidth}{!} {
		
		\begin{tabular}{c|c|llr|llr|llr|llr|llr|llr}
			\toprule
			\multirow{3}{*}{\textbf{MR}} & \multirow{3}{*}{\textbf{Feat$^\perp$}} &\multicolumn{18}{c}{\textbf{Performance measurements}}\\
			&&\multicolumn{3}{c|}{\textbf{Accuracy}}
			 &\multicolumn{3}{c|}{\textbf{Precision}}			&\multicolumn{3}{c|}{\textbf{Recall}}
			 &\multicolumn{3}{c|}{\textbf{f-measure}}
			 &\multicolumn{3}{c|}{\textbf{AUC}}
			 &\multicolumn{3}{c}{\textbf{BSR}}\\
			&& $DS_{JK}$& $DS_{JV}$	& Error$^\pm$ & $DS_{JK}$ & $DS_{JV}$ & Error$^\pm$
			 & $DS_{JK}$& $DS_{JV}$	& Error$^\pm$ & $DS_{JK}$ & $DS_{JV}$ & Error$^\pm$
			 & $DS_{JK}$& $DS_{JV}$	& Error$^\pm$ & $DS_{JK}$ & $DS_{JV}$ & Error$^\pm$\\
			\midrule
			\multirow{3}{*}{ADD}&
			 NF-NP	
			& \textbf{0.802}& 0.787	& 0.015	& \textbf{0.786}& 0.751	& 0.035	& \textbf{0.812}	& 0.704	& 0.108	
			& 0.773	&\textbf{0.775}	&\textbf{-0.002}	& \textbf{0.837}& 0.827	& 0.010	& 0.768	& \textbf{0.785}& \textbf{-0.017} \\
			
			& GK	
			& 0.712	& \textbf{0.816}& \textbf{-0.104}& 0.702 & \textbf{0.732}& \textbf{-0.030}	& 0.717	& \textbf{0.758}&\textbf{ -0.041}
			& \textbf{0.744} & 0.712& 0.032	& \textbf{0.769} & 0.707	& 0.062	&\textbf{0.737}& 0.729	& 0.008	\\
			
			& RWK 
			& 0.851	& \textbf{0.86}	& \textbf{-0.009}& \textbf{0.836} & 0.712 & 0.124 & 0.771 & \textbf{0.791} & \textbf{-0.020}	
			& \textbf{0.786} & 0.785 & 0.001 & \textbf{0.905} & 0.877 & 0.028 & \textbf{0.843}& 0.829 & 0.014	\\

			\midrule
			\multirow{3}{*}{MUL}&
			NF-NP
			& \textbf{0.712}& 0.688	& 0.024	& 0.672	& \textbf{0.689} & \textbf{-0.017} & \textbf{0.685} & 0.661 & 0.024
			& 0.657	& \textbf{0.705}& \textbf{-0.048}& \textbf{0.742} & 0.734	& 0.008	 & 0.631 & \textbf{0.654} & \textbf{-0.023} \\
			
			& GK
			&\textbf{0.663}	& 0.641	& 0.022	& 0.714	& \textbf{0.732} & \textbf{-0.018} & 0.697 & \textbf{0.758} & \textbf{-0.061}
			&0.676 & \textbf{0.733} & \textbf{-0.057} & \textbf{0.775} & 0.730 & 0.045 &	\textbf{0.689}& 0.657 &	0.032 \\
			
			& RWK 
			& \textbf{0.789}& 0.695	& 0.094	& 0.666	& \textbf{0.706} & \textbf{-0.040} & 0.693 & \textbf{0.797} & \textbf{-0.104}
			& 0.660	& \textbf{0.676} & \textbf{-0.016} & \textbf{0.846}	& 0.820	& 0.026	& \textbf{0.774} & 0.739 & 0.035	\\

			\midrule
			\multirow{3}{*}{PER}&
			NF-NP 
			& 0.838	& \textbf{0.840} & \textbf{-0.002} & 0.860 & \textbf{0.883} & \textbf{-0.023} & 0.835 & \textbf{0.846} & \textbf{-0.011}
			& \textbf{0.855} & 0.790 & 0.065	& \textbf{0.945} & 0.925 & 0.020	& \textbf{0.847} & 0.813 & 0.034	\\
			
			& GK 
			& \textbf{0.834}& 0.826	& 0.008	& \textbf{0.888}& 0.819	& 0.069	& 0.823& \textbf{0.845}	&\textbf{-0.022}	
			& 0.864	& 0.79	& 0.074	& \textbf{0.872} & 0.811 & 0.061 & \textbf{0.853}& 0.839	& 0.014	\\
			
            & RWK
            & 0.916	& \textbf{0.918} & \textbf{-0.002} & \textbf{0.917} & 0.827	& 0.090	& \textbf{0.878} & 0.835 & 0.043	
            & 0.877	& \textbf{0.893} & \textbf{-0.016} & \textbf{0.963}	& 0.944	& 0.019	& 0.757	& \textbf{0.793} & \textbf{-0.036} \\

            \midrule
			\multirow{3}{*}{INC}&
            PF-NP
            & 0.792	& \textbf{0.807} & \textbf{-0.015} & \textbf{0.847} & 0.822 & 0.025 & \textbf{0.837} & \textbf{0.837} &	\textbf{0.000} 
            & \textbf{0.776} & 0.759 & 0.017	 & \textbf{0.845} & 0.852 & \textbf{-0.007}& \textbf{0.793} & 0.786 &	0.007\\

            & GK
            & 0.752	& \textbf{0.788} & \textbf{-0.036} & 0.721 & \textbf{0.781} & \textbf{-0.060} & \textbf{0.790} & 0.776 & 0.014
            & \textbf{0.783} & 0.718 & 0.065 & 0.850  & \textbf{0.882} & \textbf{-0.032} & \textbf{0.762} & 0.744 & 0.018 \\
            
            & RWK
            & 0.799	& \textbf{0.839} & \textbf{-0.040} & \textbf{0.832} & 0.792 & 0.040 & \textbf{0.800} & 0.773 & 0.027
            & \textbf{0.854} & 0.764 & 0.090 & \textbf{0.862} & 0.821 & 0.041 &	\textbf{0.673}& 0.654 & 0.019 \\

            \midrule
			\multirow{3}{*}{EXC}&
			NF-NP
			& 0\textbf{.763}& 0.753	& 0.010	& 0.772	& \textbf{0.783} & \textbf{-0.011} & \textbf{0.778} & 0.759 & 0.019 
			& 0.762	& \textbf{0.789} & \textbf{-0.027} & \textbf{0.768} & 0.755	& 0.013& \textbf{0.868} & 0.839 & 0.029\\
			
			& GK
			& \textbf{0.816}& 0.787	& 0.029	& \textbf{0.816} & 0.861 & \textbf{-0.045} & 0.849 & \textbf{0.890} & \textbf{-0.041}
			& \textbf{0.871}& 0.790	& 0.081	& \textbf{0.873} & 0.870	& 0.003	 & \textbf{0.758} & 0.755 & 0.003	\\
			
            & RWK
            & \textbf{0.774}& 0.725	& 0.049	& \textbf{0.757}& 0.743	& 0.014	& \textbf{0.757} & 0.741 & 0.016	
            & \textbf{0.769}& 0.744	& 0.025	& \textbf{0.731}& 0.727	& 0.004	& \textbf{0.79}	& 0.757	& 0.033	\\

            \midrule
			\multirow{3}{*}{INV}&
            NF-NP 
            & \textbf{0.714} & 0.705 & 0.009 & \textbf{0.674} & 0.659 & 0.015 & \textbf{0.702} & 0.671	& 0.031	
            & 0.675	& \textbf{0.694} & \textbf{-0.019}& 0.905 & \textbf{0.917} & \textbf{-0.012} & 0.656	& \textbf{0.661} & \textbf{-0.005} \\
            
            &GK
            &\textbf{0.778}	& 0.759	& 0.019	& 0.765	& \textbf{0.769	}& \textbf{-0.004}& \textbf{0.738}	& 0.737	& 0.001	
            & \textbf{0.760}	& 0.721	& 0.039	& \textbf{0.671}	& 0.670	& 0.001	& \textbf{0.679}	& 0.655	& 0.024	\\
            
            &RWK	
            & \textbf{0.651}	& 0.585	& 0.066	& 0.610	& \textbf{0.659} & \textbf{-0.049}& \textbf{0.643} & 0.639	& 0.004	
            & \textbf{0.675}	& 0.653	& 0.022	& 0.760	& \textbf{0.766} & \textbf{-0.006}& \textbf{0.787} & 0.639	& 0.148	\\

			\bottomrule
			\multicolumn{18}{l}{{$^\perp$Feature extraction approach,  $^\pm$Error ($DS_{JK}-DS_{JV}$), \textbf{NF-PF:} Node Feature - Path Feature}, \textbf{GK:} Graphnet Kernel, \textbf{RWK:} Random Walk Kernel } 
	\end{tabular}}
\end{table*}

%% file: RQ1.tex
\extrarowheight = 0.25ex
\begin{table}[!ht]
	\caption{Comparison of PMR performance (AUC and BSR) achieved by \citeauthor{PMR3} and when using classifiers developed by us starting from $DS_{JK}$}
	\label{tbl:RQ1}
	\centering
	\resizebox{0.48\textwidth}{!} {
		
		\begin{tabular}{c|c|llr|llr}
			\toprule
			\multirow{3}{*}{\textbf{MR}} & \multirow{3}{*}{\textbf{Feat$^\perp$}} &\multicolumn{6}{c}{\textbf{Performance measurements}}\\
			&&\multicolumn{3}{c|}{\textbf{AUC}}
			 &\multicolumn{3}{c}{\textbf{BSR}} \\
			&& \cite{PMR3} & $DS_{JK}$& Error$^\pm$ & \cite{PMR3} & $DS_{JK}$& Error$^\pm$ \\
			\midrule
			\multirow{3}{*}{ADD}
			 &NF-PF	&	0.81	&	\textbf{0.837}	&	\textbf{-0.027}	&	\textbf{0.77}	&	0.768	&	0.002	\\
             &GK	&	\textbf{0.83}	&	0.769	&	0.061	&	\textbf{0.79}	& 0.737	&	0.053	\\
             &	RWK	&	\textbf{0.92}	&	0.905	&	0.015	&	\textbf{0.85}	&	0.843	&	0.007	\\

			\midrule
			\multirow{3}{*}{MUL}
			&	NF-PF	&	0.73	&	\textbf{0.742}	&	\textbf{-0.012}	&	0.65	&	0.631	&	0.019	\\
            &	GK	&	\textbf{0.78}	&	0.775	&	0.005	&	\textbf{0.69}	&	0.689	&	0.001	\\
            &	RWK	&	0.83	&	\textbf{0.846}	&	\textbf{-0.016}	&	0.74	&	\textbf{0.774}	&	\textbf{-0.034}	\\

			\midrule
			\multirow{3}{*}{PER}
			&	NF-PF	&	0.93	&	\textbf{0.945}	&	\textbf{-0.015}	&	0.83	&	\textbf{0.847}	&	\textbf{-0.017}	\\
            &	GK	&	\textbf{0.91}	&	0.872	&	0.038	&	0.83	&	\textbf{0.853}	&	\textbf{-0.023}	\\
            &	RWK	&	0.95	&	\textbf{0.963}	&	\textbf{-0.013}	&	\textbf{0.87}	&	0.757	&	0.113	\\

            \midrule
			\multirow{3}{*}{INC}
            &	NF-PF	&	0.84	&	\textbf{0.845}	&	\textbf{-0.005}	&	\textbf{0.80}	&	0.793	&	0.007	\\
            &	GK	&	\textbf{0.88}	&	0.850	&	0.030	&	0.75	&	\textbf{0.762}	&	\textbf{-0.012}	\\
            &	RWK	&	\textbf{0.89}	&	0.862	&	0.028	&	\textbf{0.79}	&	0.673	&	0.117	\\

            \midrule
			\multirow{3}{*}{EXC}
            &	NF-PF&	\textbf{0.78}	&	0.768	&	0.012	&	0.75	&	\textbf{0.868}	&	\textbf{-0.118}	\\
            &	GK	&	0.78	&	\textbf{0.873}	&	\textbf{-0.093}	&	0.74	&	\textbf{0.758}	&	\textbf{-0.018}	\\
            &	RWK	&	\textbf{0.90}	&	0.731	&	0.169	&	\textbf{0.79}	&	\textbf{0.790}	&	\textbf{0.000}	\\

            \midrule
			\multirow{3}{*}{INV}
			&	NF-PF	&	0.84	&	\textbf{0.905}	&	\textbf{-0.065}	&	0.64	&	\textbf{0.656}	&	\textbf{-0.016}	\\
            &	GK	&	\textbf{0.68}	&	0.671	&	0.009	&	0.66	&	\textbf{0.679}	&	\textbf{-0.019}	\\
            &	RWK	&	0.76	&	\textbf{0.769}	&	\textbf{-0.009}	&	0.74	&	\textbf{0.787}	&	\textbf{-0.047}	\\

			\bottomrule
			\multicolumn{8}{l}{$^\perp$Feature extraction approach,  $^\pm$Error (~\cite{PMR3}$-DS_{JK}$), \cite{PMR3}: \citeauthor{PMR3} }\\ \multicolumn{8}{l}{\textbf{NF-PF:} Node and Path Feature, \textbf{GK:} Graphnet Kernel} \\
			\multicolumn{8}{l}{\textbf{RWK:} Random Walk Kernel}
	\end{tabular}}
\end{table}

%% file: RQ2_PY-C++.tex
\extrarowheight = 0.25ex
\begin{table*}[!ht]
	\caption{Performance of SVM models when trained with $DS_{JV}$ and tested with $DS_{PY}$ and $DS_{C++}$}
	\label{tbl:RQ2_PY-C++}
	\centering
	\resizebox{\textwidth}{!} {
		
		\begin{tabular}{c|c|ll|ll|ll|ll|ll|ll}
			\toprule
			\multirow{3}{*}{\textbf{MR}} & \multirow{3}{*}{\textbf{Feat$^\perp$}} &\multicolumn{12}{c}{\textbf{Performance measurements}}\\
			&&\multicolumn{2}{c|}{\textbf{Accuracy}}
			 &\multicolumn{2}{c|}{\textbf{Precision}}			&\multicolumn{2}{c|}{\textbf{Recall}}
			 &\multicolumn{2}{c|}{\textbf{f-measure}}
			 &\multicolumn{2}{c|}{\textbf{AUC}}
			 &\multicolumn{2}{c}{\textbf{BSR}}\\
			&& $DS_{PY}$ & $DS_{C++}$ & $DS_{PY}$ & $DS_{C++}$  
			 & $DS_{PY}$ & $DS_{C++}$ & $DS_{PY}$ & $DS_{C++}$ 
			 & $DS_{PY}$ & $DS_{C++}$ & $DS_{PY}$ & $DS_{C++}$\\
			\midrule
			\multirow{3}{*}{ADD}
			&NF-PF	&	0.575	&	0.459	&	0.572	&	0.522	&	0.555	&	0.551	&	0.554	&	0.473	&	0.551	&	0.529	&	0.563	&	0.466	\\
            &GK	    &	0.564	&	0.447	&	0.532	&	0.426	&	0.543	&	0.470	&	0.531	&	0.473	&	0.547	&	0.452	&	0.561	&	0.427	\\
            &RWK	&	0.544	&	0.468	&	0.526	&	0.474	&	0.593	&	0.403	&	0.500	&	0.483	&	0.550	&	0.466	&	0.503	&	0.414	\\

			\midrule
			\multirow{3}{*}{MUL}
			&NF-PF	&	0.494	&	0.588	&	0.627	&	0.596	&	0.495	&	0.639	&	0.522	&	0.658	&	0.623	&	0.574	&	0.652	&	0.648	\\
			&GK 	&	0.460	&	0.472	&	0.463	&	0.436	&	0.477	&	0.392	&	0.479	&	0.400	&	0.480	&	0.431	&	0.475	&	0.478	\\
			&RWK	&	0.499	&	0.388	&	0.499	&	0.388	&	0.492	&	0.387	&	0.488	&	0.393	&	0.490	&	0.393	&	0.499	&	0.384	\\

			\midrule
			\multirow{3}{*}{PER}
            &NF-PF	&	0.521	&	0.445	&	0.503	&	0.403	&	0.517	&	0.411	&	0.507	&	0.570	&	0.535	&	0.519	&	0.542	&	0.545	\\
            &GK 	&	0.520	&	0.458	&	0.525	&	0.398	&	0.563	&	0.431	&	0.546	&	0.392	&	0.499	&	0.399	&	0.535	&	0.454	\\
            &RWK	&	0.515	&	0.424	&	0.515	&	0.540	&	0.503	&	0.471	&	0.517	&	0.496	&	0.516	&	0.413	&	0.532	&	0.514	\\

            \midrule
			\multirow{3}{*}{INC}
            &NF-PF	&	0.579	&	0.578	&	0.576	&	0.527	&	0.580	&	0.496	&	0.580	&	0.590	&	0.597	&	0.603	&	0.577	&	0.477	\\
            &GK	    &	0.578	&	0.518	&	0.588	&	0.501	&	0.597	&	0.576	&	0.579	&	0.476	&	0.582	&	0.558	&	0.594	&	0.587	\\
            &RWK	&	0.536	&	0.521	&	0.525	&	0.444	&	0.508	&	0.528	&	0.536	&	0.526	&	0.512	&	0.478	&	0.500	&	0.486	\\

            \midrule
			\multirow{3}{*}{EXC}
            &NF-PF	&	0.637	&	0.440	&	0.639	&	0.497	&	0.535	&	0.502	&	0.591	&	0.507	&	0.600	&	0.525	&	0.639	&	0.478	\\
            &	GK	&	0.597	&	0.561	&	0.648	&	0.506	&	0.592	&	0.552	&	0.649	&	0.494	&	0.610	&	0.467	&	0.658	&	0.492	\\
            &RWK	&	0.568	&	0.548	&	0.579	&	0.564	&	0.613	&	0.610	&	0.579	&	0.627	&	0.570	&	0.590	&	0.637	&	0.562	\\

            \midrule
			\multirow{3}{*}{INV}
			&NF-PF	&	0.531	&	0.493	&	0.534	&	0.422	&	0.525	&	0.433	&	0.502	&	0.471	&	0.514	&	0.411	&	0.512	&	0.491	\\
            &	GK	&	0.472	&	0.411	&	0.478	&	0.395	&	0.473	&	0.401	&	0.477	&	0.421	&	0.468	&	0.403	&	0.469	&	0.459	\\
            &	RWK	&	0.470	&	0.417	&	0.507	&	0.378	&	0.529	&	0.400	&	0.465	&	0.402	&	0.435	&	0.333	&	0.461	&	0.352	\\

			\bottomrule
			\multicolumn{14}{l}{{$^\perp$Feature extraction approach, \textbf{NF-PF:} Node Feature - Path Feature}, \textbf{GK:} Graphnet Kernel, \textbf{RWK:} Random Walk Kernel } 
	\end{tabular}}
\end{table*}

%% file: RQ3_PY-C++.tex
\extrarowheight = 0.25ex
\begin{table*}[!ht]
	\caption{Performance of SVM models for $DS_{PY}$ and $DS_{C++}$ datasets}
	\label{tbl:RQ3_PY-C++}
	\centering
	\resizebox{\textwidth}{!} {
		
		\begin{tabular}{c|c|ll|ll|ll|ll|ll|ll}
			\toprule
			\multirow{3}{*}{\textbf{MR}} & \multirow{3}{*}{\textbf{Feat$^\perp$}} &\multicolumn{12}{c}{\textbf{Performance measurements}}\\
			&&\multicolumn{2}{c|}{\textbf{Accuracy}}
			 &\multicolumn{2}{c|}{\textbf{Precision}}			&\multicolumn{2}{c|}{\textbf{Recall}}
			 &\multicolumn{2}{c|}{\textbf{f-measure}}
			 &\multicolumn{2}{c|}{\textbf{AUC}}
			 &\multicolumn{2}{c}{\textbf{BSR}}\\
			&& $DS_{PY}$ & $DS_{C++}$ & $DS_{PY}$ & $DS_{C++}$  
			 & $DS_{PY}$ & $DS_{C++}$ & $DS_{PY}$ & $DS_{C++}$ 
			 & $DS_{PY}$ & $DS_{C++}$ & $DS_{PY}$ & $DS_{C++}$\\
			\midrule
			\multirow{3}{*}{ADD}
			&NF-PF	&	0.706	&	0.577	&	0.742	&	0.660	&	0.748	&	0.590	&	0.683	&	0.645	&	0.723	&	0.691	&	0.760	&	0.653	\\
            &GK	    &	0.724	&	0.599	&	0.671	&	0.708	&	0.713	&	0.655	&	0.757	&	0.606	&	0.730	&	0.652	&	0.798	&	0.667	\\
            &RWK	&	0.737	&	0.738	&	0.653	&	0.720	&	0.699	&	0.797	&	0.668	&	0.730	&	0.693	&	0.750	&	0.726	&	0.725	\\

			\midrule
			\multirow{3}{*}{MUL}
            &NF-PF	&	0.670	&	0.611	&	0.652	&	0.623	&	0.701	&	0.584	&	0.787	&	0.688	&	0.746	&	0.594	&	0.656	&	0.580	\\
            &GK	&	0.613	&	0.658	&	0.627	&	0.657	&	0.659	&	0.648	&	0.643	&	0.561	&	0.663	&	0.607	&	0.663	&	0.625	\\
            &RWK	&	0.727	&	0.795	&	0.732	&	0.742	&	0.640	&	0.685	&	0.726	&	0.715	&	0.686	&	0.735	&	0.721	&	0.677	\\

			\midrule
			\multirow{3}{*}{PER}
            &NF-PF	&	0.818	&	0.732	&	0.822	&	0.702	&	0.820	&	0.778	&	0.755	&	0.763	&	0.769	&	0.754	&	0.865	&	0.754	\\
            &GK	&	0.835	&	0.777	&	0.785	&	0.725	&	0.820	&	0.830	&	0.802	&	0.752	&	0.797	&	0.717	&	0.829	&	0.694	\\
            &RWK	&	0.869	&	0.824	&	0.856	&	0.853	&	0.811	&	0.877	&	0.862	&	0.755	&	0.796	&	0.735	&	0.798	&	0.840	\\

            \midrule
			\multirow{3}{*}{INC}
            &NF-PF	&	0.746	&	0.677	&	0.734	&	0.684	&	0.789	&	0.661	&	0.796	&	0.705	&	0.789	&	0.647	&	0.792	&	0.619	\\
            &GK	    &	0.671	&	0.754	&	0.659	&	0.713	&	0.685	&	0.756	&	0.685	&	0.781	&	0.682	&	0.772	&	0.681	&	0.672	\\
            &RWK	&	0.785	&	0.760	&	0.793	&	0.721	&	0.808	&	0.784	&	0.746	&	0.682	&	0.804	&	0.753	&	0.793	&	0.694	\\

            \midrule
			\multirow{3}{*}{EXC}
            &NF-PF	&	0.734	&	0.635	&	0.694	&	0.649	&	0.713	&	0.652	&	0.725	&	0.690	&	0.715	&	0.732	&	0.737	&	0.732	\\
            &GK	    &	0.752	&	0.698	&	0.735	&	0.681	&	0.703	&	0.745	&	0.725	&	0.742	&	0.764	&	0.760	&	0.734	&	0.673	\\
            &RWK	&	0.791	&	0.805	&	0.805	&	0.834	&	0.782	&	0.794	&	0.788	&	0.786	&	0.808	&	0.811	&	0.780	&	0.753	\\

            \midrule
			\multirow{3}{*}{INV}
			&NF-PF	&	0.606	&	0.571	&	0.671	&	0.543	&	0.668	&	0.539	&	0.661	&	0.594	&	0.673	&	0.559	&	0.672	&	0.538	\\
            &GK	&	0.582	&	0.595	&	0.598	&	0.620	&	0.589	&	0.586	&	0.558	&	0.579	&	0.604	&	0.633	&	0.599	&	0.624	\\
            &RWK	&	0.683	&	0.611	&	0.673	&	0.629	&	0.717	&	0.614	&	0.648	&	0.649	&	0.675	&	0.711	&	0.712	&	0.708	\\

			\bottomrule
			\multicolumn{14}{l}{{$^\perp$Feature extraction approach, \textbf{NF-PF:} Node Feature - Path Feature}, \textbf{GK:} Graphnet Kernel, \textbf{RWK:} Random Walk Kernel } 
	\end{tabular}}
\end{table*}